# Phase transitions in a Kagomé lattice of Josephson junctions

*M.S. Rzchowski*

*Physics Department and Applied Superconductivity Center*

*University of Wisconsin-Madison*

*Madison, WI 53706*

We investigate the nature of the phase transition in Josephson junctions arranged on a Kagomé lattice. We find that an applied magnetic field corresponding to 1/2 flux quanta per elementary triangle results in a $\pi$ phase shift in the current phase relation for *all* bonds, resulting in an *XY* antiferromagnet. This corresponds to the order-from-disorder selected highly-degenerate coplanar state of the more extensively studied Heisenberg Kagomé antiferromagnet. Using an histogram Monte-Carlo analysis, we observe a phase transition at 0.078 of the coupling energy. We find that the jump in the superfluid density at the transition temperature, determined from a finite-size scaling analysis of the magnetization fluctuations, retains its universal value within the Kosterlitz-Thouless scenario despite the well-known degeneracy of the low-temperature phase. This universal jump, combined with the dramatically suppressed transition temperature, point to a very strong renormalization of the spin stiffness just below the transition temperature. We also observe a sequence of ground states corresponding to flux quanta per elementary triangle of 1/2, 3/8, 1/4, 1/8, and 0 that, unlike ground states of other lattices, consist of currents of equal magnitude circulating around each elementary triangle in the structure. This sequence of structurally similar ground states will permit a detailed analysis of the dependence of transition temperature on frustration.

1995 PACS numbers: 75.10.Hk, 74.50.+r, 75.40.Mg

The Heisenberg Kagomé antiferromagnet has gained recent attention in its experimentally accessible incarnation $SrCr_{8-x}Ga_{4+x}O_{19}$, which has been found to have two-dimensional (2D) Kagomé planes (Fig. 1) of antiferromagnetically-coupled $Cr^{3+}$ magnetic ions, separated by more diluted triangular planes[1]. This system of magnetic spins on the Kagomé planes has the distinguishing factor of allowing a classical ground state "entropy" at zero temperature that scales with the number of sites in the lattice[2]. At temperatures of the order of $10^{-2}J / k$, where $J$ is the coupling energy and $k$ Boltzmann's constant, this 2D Heisenberg Kagomé system is known to select out states with the Heisenberg spins lying in the plane of the lattice *via* an order-from-disorder mechanism[3]. At a temperature below this, there is theoretical evidence that quantum and thermal fluctuations further select one of the planar states over the others[4].

Arrays of Josephson junctions provide experimentally accessible and controllable systems in which to explore phase transitions in 2D classical magnets. Experimental systems consist of photolithographically defined 2D arrays of superconducting "islands" coupled weakly to their nearest neighbors by either a normal metal[5] or insulating[6] barrier. The Josephson junctions thus formed have an energy in zero applied magnetic field proportional to the cosine of the difference of the phase of the superconducting order parameter on the two islands coupled by the junction. The magnetic analogy is then made by considering the phase of each island to be the angle that the corresponding XY spin makes with a defined axis. This system has been used to experimentally and theoretically study[7] the Kosterlitz-Thouless (KT) transition in XY magnets.

An external magnetic field applied perpendicular to the sample does not couple directly to "spins" in the Josephson system as it does in a magnet, but instead introduces into the expression for the coupling energy a phase shift proportional to the line integral of the vector potential between lattice sites, resulting in a Hamiltonian for the system that can be written[8]

$$H = -\sum_{i,j \in \{n.n.\}} E_J \cos\left(\varphi_i - \varphi_j - A_{ij}\right) \tag{1}$$

where

$$A_{ij} = \left(2\pi / \Phi_o\right)\int_i^j \vec{A} \cdot d\ell$$

and $\varphi_i$ is the phase of the $i^{th}$ superconducting island, $\vec{A}$ is the vector potential representing the applied magnetic field, $E_J$ the Josephson coupling energy $\hbar I_C / 2e$, with $I_C$ the critical current of the junction, and $\Phi_o$ is the flux quantum $hc / 2e$. The sum is restricted to nearest neighbors. In zero applied magnetic field the effective coupling is ferromagnetic, and results in a Kosterlitz-Thouless phase transition to a low temperature state with power-law correlations. For nonzero perpendicular field, the Hamiltonian of the system and the associated ground states can be quite rich[8] due to the additional periodicity of the Hamiltonian introduced by the vector potential competing with the periodicity of the lattice. However it is clear that bonds with an energy expression involving a $\pi$ phase shift can be considered to connect two antiferromagnetically-coupled spins. In many lattices, it is not possible for all bonds to be simultaneously antiferromagnetic. We have found, however, that an applied magnetic field producing a flux of $\Phi_o / 2$ through an elementary triangle of the Kagomé junction array leads to an antiferromagnetic character for all bonds[9], and hence a thermodynamically significant ground state entropy[2]. In this paper, we investigate *via* Monte-Carlo simulations the effects of such a ground-state entropy on the thermodynamic phase transition. We also discuss a particular series of five magnetic fields corresponding to flux quanta per elementary triangle $f$, also termed frustration, between 1/2 and 0 that result in a sequence of ground states identical apart from a scaling factor, rather than the accommodation of an Abrikosov vortex lattice to the junction array observed in all previously investigated lattices[10].

We have carried out Monte-Carlo simulations of Josephson junctions arranged on a Kagomé lattice (Fig. 1) defined by the Hamiltonian (1) for magnetic fields corresponding to $f$=0, 1/8, 1/4, 3/8, and 1/2 and lattices containing up to 432 sites with periodic boundary conditions. We have determined the ground states for these magnetic fields by cooling from the disordered high-temperature phase to a temperature at which the spin structure was stable as a function of Monte Carlo time step, up to an overall rotation of each spin, for which the Hamiltonian (1) is invariant.

The ground state as illustrated by the angle of each spin can be quite complicated, but in terms of bond variables, or gauge-invariant phase differences[11], $\gamma_{ij} \equiv \varphi_i - \varphi_j - A_{ij}$, we have found that the ground states for the fields corresponding to $f$=0, 1/8, 1/4, 3/8, and 1/2 are surprisingly characterized by all bond variables having equal magnitude. Since the Josephson supercurrent between two superconducting islands is equal to the sine of the bond variable[11], we can display the nondegenerate ground states for $f$ = 0, 1/8, 1/4, and 3/8 as in Fig. 1, with the direction of the arrow representing the direction of the supercurrent. The ground states for $f$ = 1/2 correspond to antiferromagnetic coupling and reflects the known degeneracy in that the chirality, or vorticity, of each elementary triangle can be either positive or negative in addition to the state shown in Fig. 1, with additional constraints regarding the arrangement[3]. The magnitudes of the bond variables for $f$=1/2 remain all equal. The bond variables take on magnitudes of 0, $\pi$/12, $\pi$/6, $\pi$/4, and $\pi$/3 for $f$ = 0, 1/8, 1/4, 3/8, and 1/2 respectively. This sequence of ground states related by a scaling factor is unique to the Kagomé lattice: ground states of previously studied lattices[10] show no such scaling. This scaling permits an unambiguous evaluation of the field dependence of the KT transition in Josephson-junction arrays[12].

This description of the ground states, with bond variables of equal magnitudes, makes a mean-field calculation of the field-dependent phase transition temperature straightforward. Associating a complex order parameter $\eta_j = \left\langle e^{i\varphi_j} \right\rangle$ with each spin in the array, where $\langle \cdots \rangle$ denotes a thermal average, the linearized mean-field equations can be written[13]

$$\eta_i = \frac{\beta E_J}{2} \sum_{j \in \{n.n\}} e^{iA_{ij}} \eta_j \tag{2}$$

a relation equivalent to the tight-binding Schrodinger equation of an electron on a lattice in uniform applied magnetic field[14]. Since the ground states are non-degenerate (apart from $f$=1/2), we expect the phase of the order parameter at the $i^{th}$ site to retain its zero-temperature value, but to have a temperature-dependent magnitude. Given the equality of the zero-temperature bond variables, the magnitude $|\eta_i|$ must be independent of lattice site. Equation (2) can then be solved for the mean-field critical temperature as $kT_c / E_J = 2\cos\gamma$ where $\gamma$ is the magnitude of the (position independent) gauge-invariant phase difference discussed above for each ground state. These transition temperatures are in agreement with the mean-field results of Lin and Nori[15] derived with a different technique, and are shown in Fig. 2 along with transition temperatures determined from our Monte Carlo simulations to be discussed later.

We have also investigated the phase transition in these applied fields with Monte Carlo simulations of lattices containing up to 432 sites with periodic boundary conditions. For all fields, we find a peak in the heat capacity per site (not shown) that saturates as a function of lattice size, characteristic of a Kosterlitz-Thouless phase transition. The *xx* component of the spin stiffness tensor, with *x* along the horizontal direction of Fig. 1, has been computed using the methods of [16]. The spin stiffness is the curvature of the free energy with respect to the wavevector of a long-wavelength spin wave, and becomes non-zero only in the ordered state. In a Kosterlitz-Thouless phase transition, the temperature-dependent spin stiffness $\Gamma(T)$ has a discontinuous jump at the transition temperature from zero to a universal[17] value of $2kT_c / \pi$. The spin stiffness normalized to the Josephson coupling energy for the range of applied fields discussed above is shown in Fig. 3, along with a line of slope $2k / \pi E_J$. Due to finite-size effects[18] no discontinuous jump is observed, although the temperature at which a non-zero spin stiffness is first obtained follows the same trend as the mean-field transition temperatures of Fig. 2.

We exploit finite size scaling properties of Binder's fourth-order cumulant of the magnetization[19], $1 - \langle m^4 \rangle / 3\langle m^2 \rangle^2$, to determine the transition temperature. This cumulant has been shown[19] within the framework of finite-size scaling to approach a universal constant, independent of lattice size, when spin correlations in the system have a power law dependence on the spatial coordinate. Thus cumulants for different system sizes would cross at the critical point of a second-order phase transition. Since correlations are power law at all temperatures below the critical point in a Kosterlitz-Thouless transition, the cumulant curves merge together below the critical temperature[20]. Cumulants for a field corresponding to a frustration of *f*=1/2 for lattices of 48, 108, 192, and 300 sites are shown in Fig. 3. The magnetization is calculated as the vector sum magnitude $\left(\left(\sum \cos(3\theta_i)\right)^2 + \left(\sum \sin(3\theta_i)\right)^2\right) / N^2$, where $\theta_i$ is the angle formed with the *x*- axis by the spin at site *i*. From this analysis we determine the *f*=1/2 transition temperature to be $0.078E_J / k$. For fields corresponding to *f*=0, 1/8, 1/4, and 3/8, the ground states are non-degenerate, and the total magnetization is determined as the sum of 1, 12, 6, and 4 sublattice magnetizations respectively. The transition temperatures resulting from the associated cumulant analyses are shown in Fig. 2. Consistent with the known effects of fluctuations, the phase transition temperatures determined from our Monte Carlo simulations lie below the mean-field results. We have also observed substantial deviation from the curve drawn through the Monte Carlo data points of Fig. 2 for values of *f* corresponding to ground states with structure much more complicated than that of Fig. 1. We reiterate here that the monotonic dependence of critical temperature on the frustration *f* shown in Fig. 2 is indicative of the identical ground-state structure for these particular values of frustration.

We now focus on the case of fully antiferromagnetic bonds (*f*=1/2), addressing with Monte Carlo simulation techniques the nature of the phase transition and the degree to which it is affected by the thermodynamically significant ground state degeneracy discussed previously. We find that the low transition temperature necessitates Monte Carlo runs of $10^6$ equilibration steps and $5\text{x}10^6$ averaging steps per site to ensure equilibrium behavior. This was established by examining thermodynamic quantities as well as the symmetry of the energy histograms[21]. The data of Figs. 4, 5, and 6 are averages of six such statistically independent Monte Carlo runs except for the 300-site data, which is derived from two independent runs. We observe significant finite-size effects in both the spin stiffness (Fig. 5) and the heat capacity (not shown). The heat capacity per site was computed using the multiple histogram technique[21] and found to have a temperature dependence similar to that observed previously in Kosterlitz-Thouless phase transitions. The maximum in the heat capacity per site is observed to quickly saturate as a function of the number of sites to a value of $1.32k$, somewhat lower than the value of $1.45k$ obtained recently[22] for a ferromagnetically-coupled square lattice where the validity of the Kosterlitz-Thouless theory has been generally accepted. The temperature of the heat capacity peak extrapolates as a function of lattice size to $0.095E_J / k$, a factor of 1.2 higher than the transition temperature determined from the cumulant analysis of Fig. 4.  It has been suggested that the heat capacity peak in the ferromagnetically-coupled XY model on a square lattice, observed[22] to occur at $1.18T_{KT}$, is to be associated with a breakdown in the dilute-gas approximation for vortex-antivortex pairs rather than a signature of a phase transition.  Although in this sense the heat capacity is not a sharp probe of the phase transition, we note that it seems not to be substantially affected by the degeneracy of the low-temperature phase.

To further investigate the antiferromagnetic bond phase transition, we have evaluated the size of the jump in the spin stiffness in order to compare with the universal value determined by Nelson and Kosterlitz[17].  This prediction is universal in the sense that coupling to perturbations such as spin-wave excitations or background potentials do not change the magnitude of the jump in the superfluid density scaled by the transition temperature.  As finite size effects prohibit us from directly measuring the magnitude of the jump, we exploit the relation[17] $\Gamma(T_C) = 2\pi\eta(T_C)kT_C$ between this jump and the exponent $\eta(T_C)$ of power-law spin-spin correlations at the critical temperature.  At a KT phase transition, the jump in the spin stiffness corresponds to an exponent $\eta$ of 0.25.  We determine the temperature-dependent value of the exponent $\eta$ from the scaling properties of the magnetization fluctuations (susceptibility) per site with the number of sites considered.  Finite-size scaling predicts[22] that the susceptibility per site should be proportional to $N^{-2\eta}$ in the temperature range where spin-spin correlations have a power law dependence, with $N$ the number of sites.  Figure 6 shows this behavior to be satisfied below the transition, while deviating from such power-law behavior at high temperature.  The interpolated value of the correlation exponent $\eta$ at the critical temperature determined from the cumulant analysis of Fig. 6 is 0.31, in fair agreement with the universal KT value of $\eta = 0.25$.  In combination with the measured critical temperature (Fig. 4), this corresponds to a normalized spin stiffness jump of only 0.05 at the transition temperature.  This jump, small in comparison with the zero temperature spin stiffness of 0.5, suggests a renormalization of the spin-stiffness below the critical temperature very much stronger than for the other applied fields of Fig. 3.

The initial decrease of the spin stiffness from its zero-temperature value on increasing temperature should be attributed to a depletion of the superfluid density by the excitation of long-wavelength spin-waves[23]. We find this behavior for $kT / E_J < 0.04$ to be accurately described by the relation $\Gamma(T) / \Gamma(0) = 1 - kT / 4E_J$ for all magnetic fields studied here. This is in agreement with the ferromagnetic-bond prediction[23] for *f*=0 ($\Gamma(0) = E_J$), but in disagreement for other values of the frustration *f*. The measured low-temperature behavior is however consistent with the non-universality expected from a classical system even at low temperatures in that the coupling energy appears in the temperature dependence rather than the zero-temperature spin-stiffness. In contrast to a quantum system, spin-wave modes out to the Brillouin zone boundary continue to be excited for arbitrarily low temperature, and hence the temperature dependence of the spin stiffness is expected to be sensitive to the lattice structure. This continuous trend for a sequence of structurally similar ground states suggests that the large low-temperature rate of depletion of superfluid density in the case of antiferromagnetic bonds is not associated with the degeneracy of the *f*=1/2 ground state, in agreement with the independence of the spin-wave excitation spectrum on the particular antiferromagnetic ground state[3]. We suggest, however, that the 90% drop in the spin stiffness before undergoing the universal jump at the critical temperature should be associated with the degeneracy of the ground state.

We have confirmed there to be no hysteresis on heating and cooling, and that the system generates thermodynamic quantities identical within our statistical error whether it is cooled from a high temperature or started in a random configuration at the temperature of interest. This suggests that the system does not become trapped in one of the degenerate ground states. It is unclear at this point why such a transition into a manifold of degenerate states should retain the universal value of the jump in the superfluid density.

We thank A. Chubukov, M. Gingras, P.C.W. Holdsworth, D. Huber, R. Joynt, and S. Teitel for useful conversations.

Figure Captions

Figure 1: Schematic representation of the ground states for *f*=0, 1/8, 1/4, 3/8, and 1/2 for an array of Josephson junctions on a Kagomé lattice. The arrows represent the direction of circulating currents. *All* currents in these ground states are of equal magnitude, unlike other lattices investigated.

Figure 2: Mean field transition temperatures (open circles) and Monte-Carlo results using a cumulant method (filled circles) as a function of flux quanta per elementary triangle *f*. The smooth trend is characteristic of the structural similarity of the ground states, as shown in Fig. 1. Zero applied field corresponds to ferromagnetic (F) bonds in the XY magnet, and *f*=1/2 corresponds to all bonds being antiferromagnetic (AF).

Figure 3: Spin stiffness normalized to the Josephson coupling energy as a function of temperature for five different values of flux quanta per elementary triangle. Data is for lattices of 432 sites with the exception of that corresponding to *f*=1/2, which is obtained from a lattice of 192 sites. Curves through the data are a guide to the eye. The line drawn has a slope of $2k / \pi E_J$, and its magnitude at the critical temperature represents the size of a universal jump in the spin stiffness at a Kosterlitz-Thouless transition temperature.

Figure 4: Binder's fourth-order magnetization cumulant as a function of temperature for various lattice sizes in a field corresponding to *f*=1/2. The critical temperature is determined as the temperature at which the curves diverge.

Figure 5: Spin stiffness normalized to the Josephson coupling energy as a function of temperature for various lattice sizes and an applied field corresponding to 1/2 flux

quanta per elementary triangle, making all bonds antiferromagnetic. This applied field results in a manifold of degenerate ground states with an entropy that scales with the number of sites.

Figure 6: Finite-size scaling analysis of magnetization fluctuations. The exponent of the power-law correlation function at the transition temperature, and hence the size of the jump in the spin stiffness, is determined from the slope of the lines.

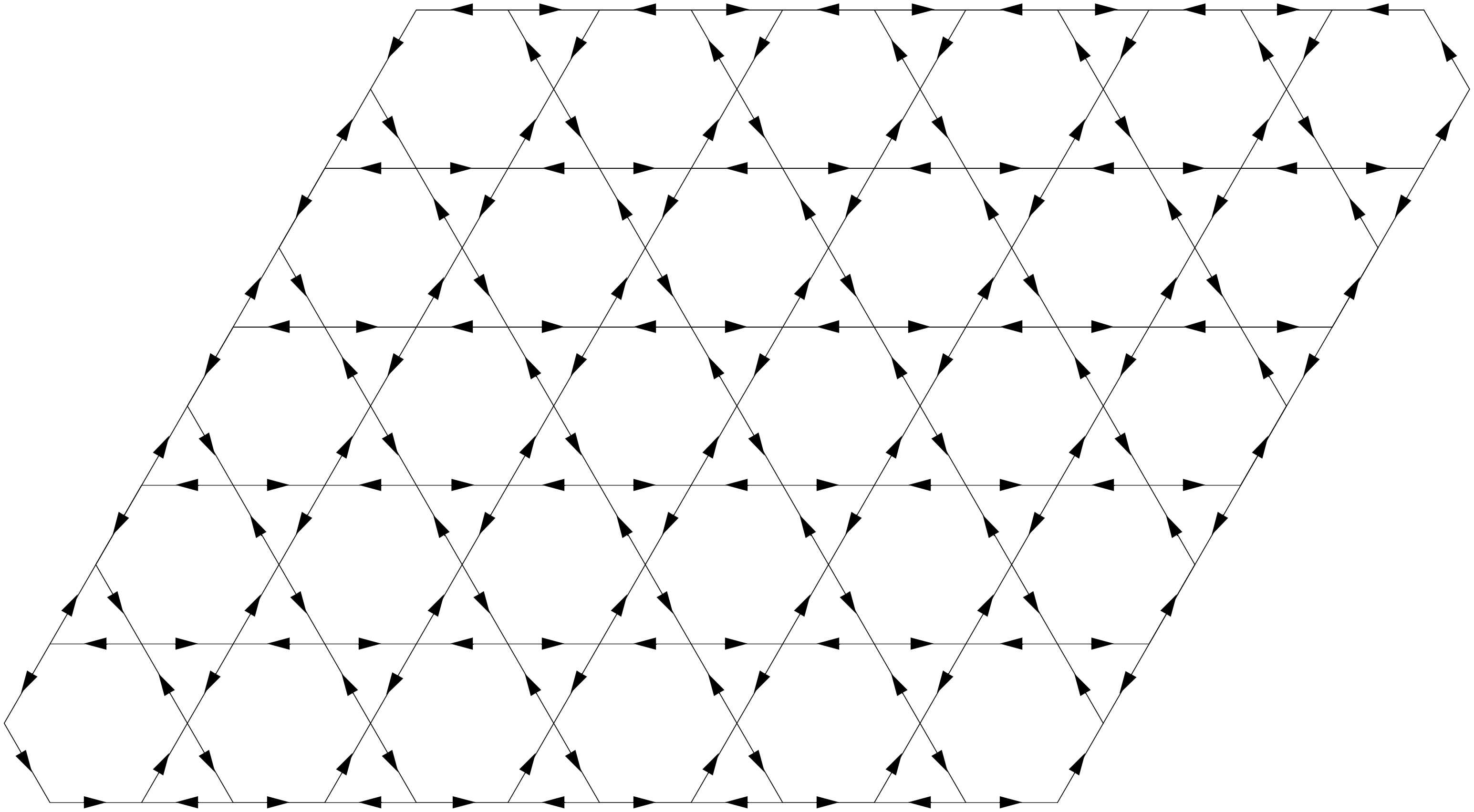

Figure 1

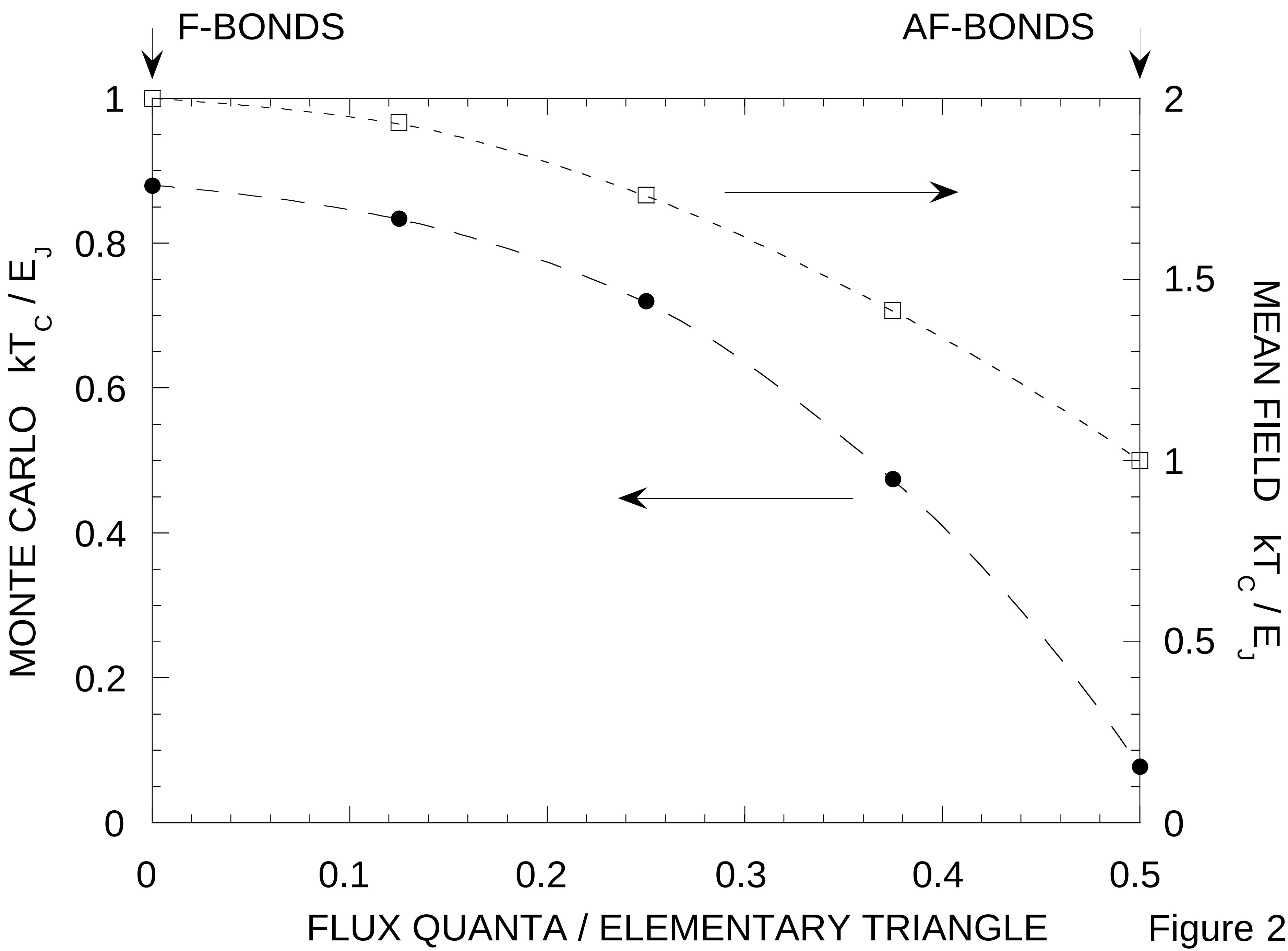

Figure 2

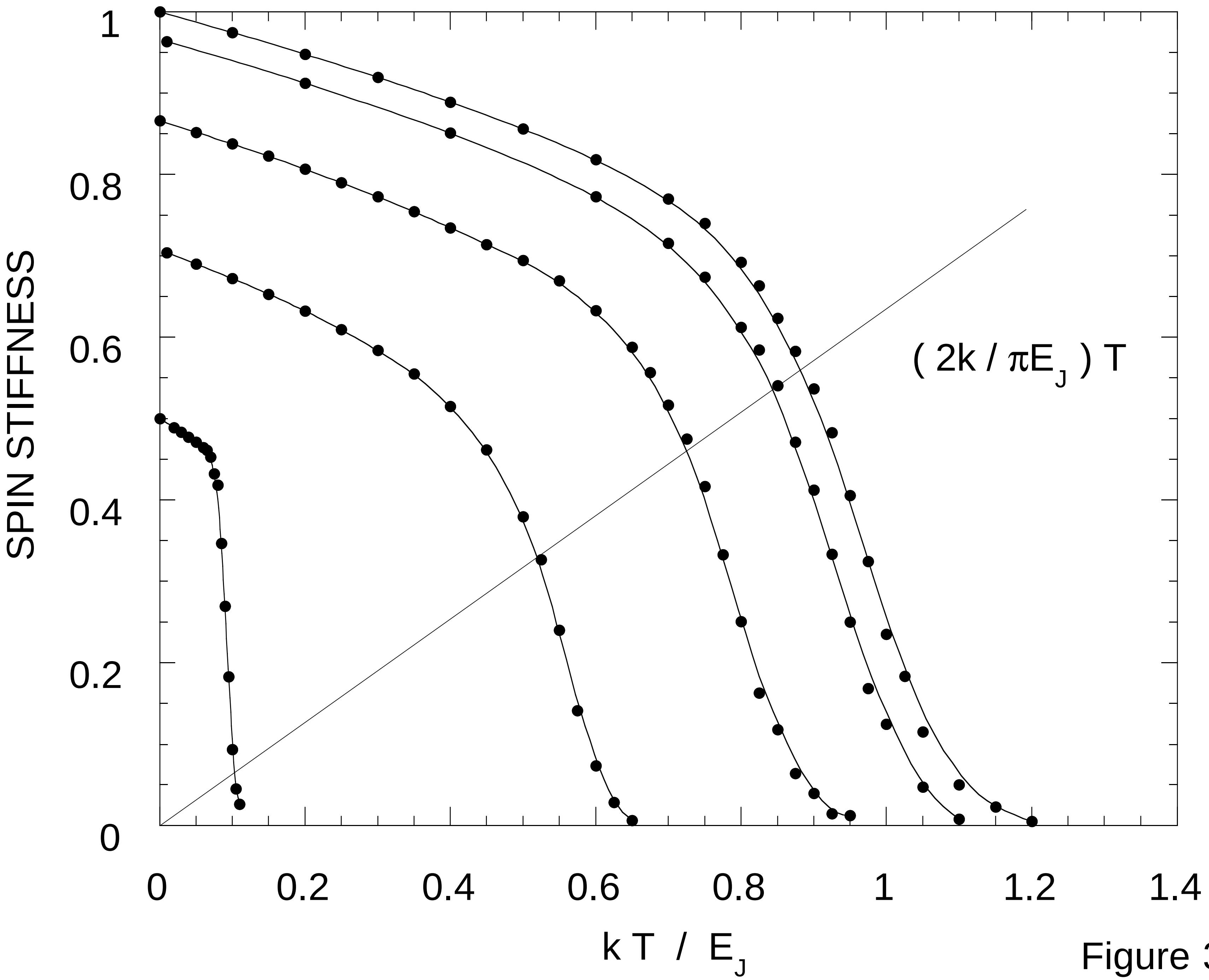

Figure 3

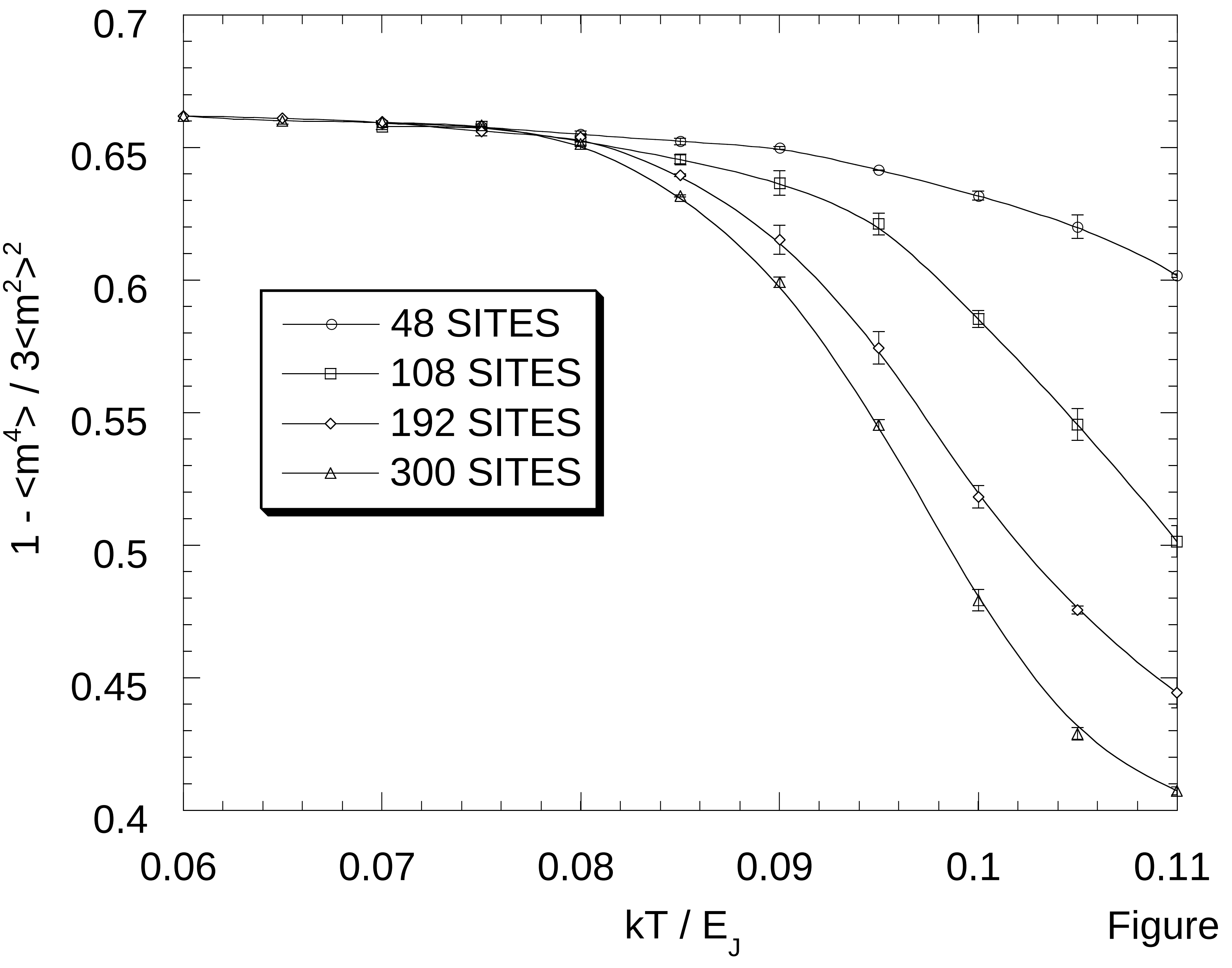

0.7
0.65
0.6
0.55
0.5
0.45
0.4
0.06
0.07
0.08
0.09
0.1
0.11
$1 - \langle m^4 \rangle / 3\langle m^2 \rangle^2$
$kT / E_J$
48 SITES
108 SITES
192 SITES
300 SITES

Figure 4

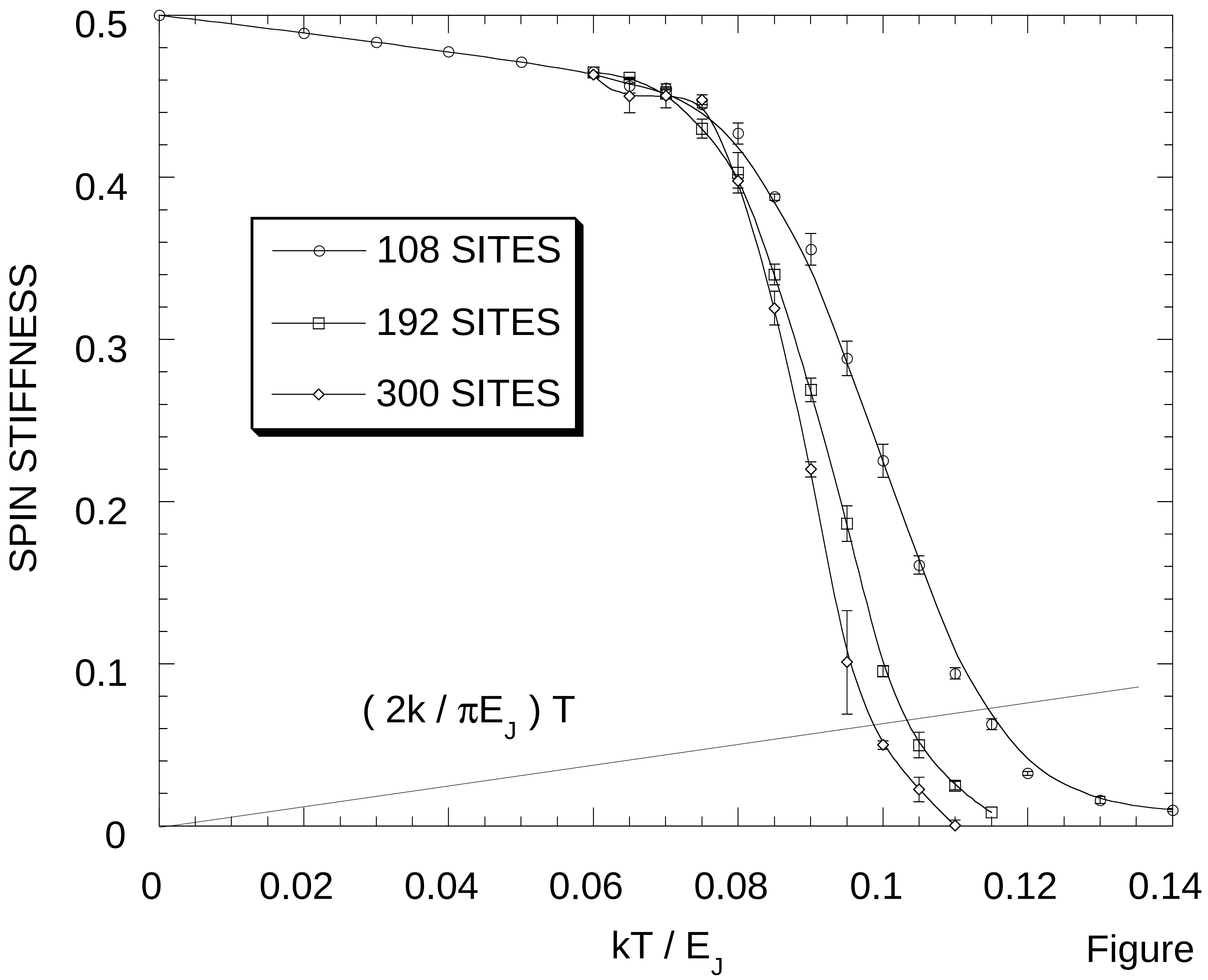

Figure 5

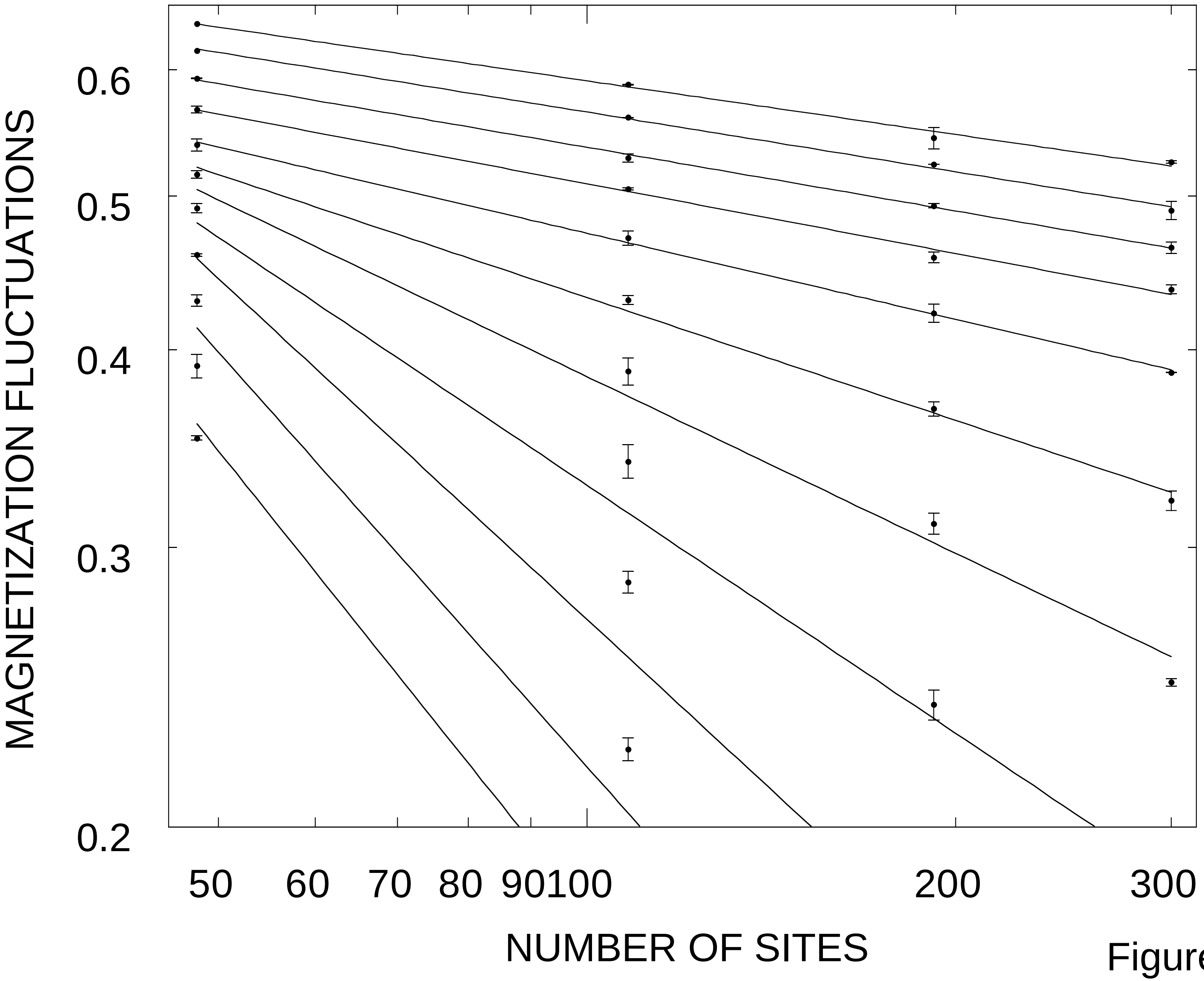

Figure 6